\documentclass[graybox]{svmult}

\usepackage{type1cm}        
\usepackage{makeidx}         
\usepackage{graphicx}        
\usepackage{multicol}        
\usepackage[bottom]{footmisc}

\usepackage{newtxtext}       %
\usepackage{newtxmath}       

\usepackage{amsfonts}
\usepackage{amsmath}

\makeindex             

\begin{document}
\begin{center}
\LARGE{ \textbf{A spectral graph theoretic study of predator-prey networks}} \\

\vspace{0.5cm}

\normalsize Shashankaditya Upadhyay$^{1}$, Sudeepto Bhattacharya$^{2}$ \\

\vspace{0.5 cm}

Department of Mathematics, School of Natural Sciences, Shiv Nadar University, Gautam Buddha Nagar, Uttar Pradesh 201 314, India \\ ${}^{1}$ Email: su348@snu.edu.in, ${}^{2}$ Email: sudeepto.bhattacharya@snu.edu.in

\vspace{1cm}
\end{center}
%
%

\abstract*{Predator-prey networks originating from different aqueous and terrestrial environments are compared to assess if the difference in environments of these networks produce any significant difference in the structure of such predator-prey networks. Spectral graph theory is used firstly to discriminate between the structure of such predator-prey networks originating from aqueous and terrestrial environments and secondly to establish that the difference observed in the structure of networks originating from these two environments are precisely due to the way edges are oriented in these networks and are not a property of random networks. We use random projections in $\mathbb{R}^{2}$ and $\mathbb{R}^{3}$ of weighted spectral distribution (WSD) of the networks belonging to the two classes viz. aqueous and terrestrial to differentiate between the structure of these networks. The spectral theory of graph non-randomness and relative non-randomness is used to establish the deviation of structure of these networks from having a topology similar to random networks. We thus establish the absence of a universal structural pattern across predator-prey networks originating from different environments.}

\abstract{Predator-prey networks originating from different aqueous and terrestrial environments are compared to assess if the difference in environments of these networks produce any significant difference in the structure of such predator-prey networks. Spectral graph theory is used firstly to discriminate between the structure of such predator-prey networks originating from aqueous and terrestrial environments and secondly to establish that the difference observed in the structure of networks originating from these two environments are precisely due to the way edges are oriented in these networks and are not a property of random networks. We use random projections in $\mathbb{R}^{2}$ and $\mathbb{R}^{3}$ of weighted spectral distribution (WSD) of the networks belonging to the two classes viz. aqueous and terrestrial to differentiate between the structure of these networks. The spectral theory of graph non-randomness and relative non-randomness is used to establish the deviation of structure of these networks from having a topology similar to random networks. We thus establish the absence of a universal structural pattern across predator-prey networks originating from different environments.}

\section{Introduction}
\label{sec:1}

Network theory has emerged in the recent years as an essential science to holistically study the interactions and relations between individual components of a complex system [1]. Networks are generic representation of complex systems in which the underlying topology is a graph. In a network, the various components of the systems are labelled as vertices and the interaction between these components is represented as edges. For a formal definition of networks, see [2]. Networks thus provide an effective approach to mathematically model empirical data from real world problems where the relationship between given components is the object of investigation. Network theory analyses the structural and functional properties of such real world network models to identify properties of the underlying complex system that may not be known previously.

Network theory had been widely used in ecology in recent years to study ecosystems and various processes originating or instantiating in these ecosystems [3,4]. Ecological networks can be broadly classified as networks of ecological connectivity of certain species [2, 6-8] and food web networks [5]. In a food web network, vertices represent species or a group of species and edges represent the relation of carbon flow between the species. In a food web network the carbon flow between the species is usually due to predation but there may exist food web networks such as host-parasite networks, plant-pollinator networks, seed-dispersal networks etc. [9-14], in which the flow of ecological information (mass-energy) is not due to predation. In this work we study food web network representations of ecological process in which the carbon flow within the vertices (species or a group of species) of the networks is mostly due to a predator-pray relationship [5,15]. In these predator-prey networks, we consider two vertices as adjacent if there is a consumption relationship between them. It must be pointed out however that non-living entities, such as detritus for instance, may also be considered a vertex in such networks.

It is generally assumed that the complexity of such predator-prey food web networks is captured in some simple algebraic measures such as connectance for example [15]. In literature the structure of these networks is often assumed to be similar to each other [16,17]. In particular, there has been no study based on spectral graph theory that attempts to distinguish between the structure of these predator-prey networks.

The primary objective of this work is to employ methods developed recently in the field of spectral graph theory to analyse if there is any considerable difference in the structure of these predator-prey networks as maybe due the difference between the ecological environment from which these networks originate. To investigate this idea, we classify some predator-prey networks studied in this work as aqueous predator-prey networks if they originate from aqueous environments or terrestrial predator-pray networks if the ecosystem they are found embedded in is a terrestrial ecosystem. In case it is found that there is no significant difference in the structure of networks originating from these two different environments, then we can assume that there is possibly a universal structural pattern in these networks and possible differences in ecological processes originating from the virtue of being embedded in different environments may play no role in producing a considerable difference in the structure of such predator-prey networks.

Spectra of a graph is often considered as a signature of the graph [18,19]. In the current study, we use the applications of weighted spectral distribution (WSD) of the normalized graph Laplacian to discriminate between the structure of predator-prey networks originating from either aqueous or terrestrial environments [20]. Random projections of weighted spectral distribution have been shown to effectively discriminate between graphs that have different topologies [21] and these spectral projections are used here to investigate if there is any difference in the structure of predator-prey networks originating from aforementioned two classes of environments.

We further try to establish that the difference which may be observed in the structure of such networks is not due to some random interactions which may be present in such networks but are because of these predator-prey networks having a different topology as compared to a random network. To quantify the deviation of a network from having a random topology, we use a spectral measure known as relative non-randomness in networks [22,23]. Relative non-randomness has been used earlier in context of social networks to quantify how different the topology of a given network is in comparison to a random network.

\section{Materials and methods} 
\label{sec:2}

A total of thirty one networks of different orders and sizes are compared for the presence of a universal structural pattern in this work. Of these thirty one networks, twenty networks originating from aqueous environments such as estuaries [24-27], lakes [28-31], seas and oceans [32-35], rivers [36-38], oxbow lakes [39] and streams [40-43] are classified as aqueous predator-prey networks.  The remaining eleven food webs are classified as terrestrial predator-prey networks as they originate from terrestrial environments such as forests [44-46], rocky shores [47-50] and other terrestrial environments [51-53] such as islands and grasslands for instance. These networks can be accessed online from a public repository of University of Canberra called GlobalWeb [\url{https://www.globalwebdb.com/}].

\subsection{Weighted spectral distribution}

The weighted spectral distribution (WSD) is a spectral measure based on the spectra of normalized graph Laplacian matrix of a graph. Given the adjacency matrix $A$ of a graph $G$, the normalized graph Laplacian $L$ of $G$ can be defined as 
\begin{equation}
L = I - D^{-\frac{1}{2}}AD^{-\frac{1}{2}}\;,
\end{equation}
where $I$ is the identity matrix and $D$ ia a diagonal matrix with entries as the degree of vertices. If $\lambda_{i} i = 0, \dots n-1$ are the eigenvalues of the normalized graph Laplacian then it is known that $0 = \lambda_{0} \leq \lambda_{1}, \dots, \leq \lambda{n-1} \leq 2$ and equality on the upper bound holds iff the graph is bipartite [18].

If we consider $K$ bins, then a function $\omega(G,N)$ on graph $G$ can be defined as:
\begin{equation}
\omega(G,N) = \sum_{k\in K} (1-k)^{N} f(\lambda = k)\;,
\end{equation}
where $N$ can be chosen as $\{2, 3, \dots\}$ and $f$ is the eigenvalue distribution of the normalized graph Laplacian of $G$.

The elements of $\omega(G,N)$ form the \textit{weighted spectral distribution} that bins the $n$ eigenvalues of the normalized graph Laplacian as:
\begin{equation}
WSD : G \Rightarrow {\mathbb{R}}^{|K|} \{ k \in K : ({(1-k)}^{N} f(\lambda = k))\}\;.
\end{equation}

The structure of a graph is related to WSD as given by the following theorem:

\begin{theorem}
The eigenvalues $\lambda_{i}$ of the normalized Laplacian matrix for an undirected network are related to the closed random walk probabilities as:
\begin{equation}
\sum_{i} {(1-\lambda_{i})}^{N} = \sum_{C} \frac{1}{d_{u_{1}} d_{u_{2}} \dots d_{u_{N}}}\;,
\end{equation}
where $N$ is the length of the random walk cycles, $d_{u_{i}}$ is the degree of vertex $u_{i}$ and $u_{1} \dots u_{N}$ denotes a closed walk from node $u_{1}$ of length $N$ ending at node $u_{N}$ such that $u_{1} = u_{N}$. Here the summation is over all possible closed walks $C$ of length $N$.
\end{theorem}

Thus the left hand side of (4) is related to WSD while the right hand side of (4) is related to distribution of small subgraphs in a graph as given by closed random walks of length $N$. For the purpose of analysis in this work, we choose $N$ as four because the corresponding WSDs in this case are related to closed random walks of length four.The closed random walk of length three are precisely the $3-cycles$ in a simple graph which are absent in bipartite graphs. Thus value of $N$ as three is not chosen for analysis.

\subsubsection{Bin selection for WSD}

Bins in WSD are assigned such that for a given value of $N$ the sum of weighting in each bin is equal. The weighting in WSD is expressed as:
\begin{equation}
w(x) = (1-x)^{N}\;,
\end{equation}
where $w(x)$ can be thought of as a function that assigns a weight to an eigenvalue of normalized graph Laplacian at $x$. The equality in the sum of weighting in each of the $K$ bins is achieved by solving the integral equation
\begin{equation}
\int_{k_{i}}^{k_{i+1}} w(x)dx = \int_{k_{j}}^{k_{j+1}} w(x)dx\;,
\end {equation}
for all $i, j$. This gives an equal weight of the function $w(x)$ in any pair of given bins $i \in (k_{i}, k_{i+1})$ and $j \in (k_{j}, k_{j+1})$.

\subsubsection{Random projections of WSD}

Random projection is a general data reduction method which is often used to reduce a high-dimensional data to low-dimensional data for the ease of computations and interpretations. Random projection of WSD has been used effectively in [19] to differentiate between the structure of graphs with different topologies. In order to distinguish $n$ graphs using WSD, consider a matrix $X \in \mathbb{R}^{n\times |K|}$ of WSDs of $n$ graphs with $K$ bins. We obtain a matrix $Y \in \mathbb{R}^{n\times d}$ by multiplying the matrix $X$ with a random projection matrix $R \in \mathbb{R}^{|K|\times d}$, where the elements of $R$ are drawn from a standard normal distribution. Thus we have
\begin{equation}
Y = XR\;,
\end{equation}
such that $R \sim N(0,1)$. The rows of $R$ in expectation form orthogonal vectors as they are normally distributed independent variables with zero correlation. Also the norm of the vectors is $1$ an thus $R$ forms a reduced basis in the original data. 

In the current study a total of thirty two networks from aqueous and terrestrial environments have been used to create a data matrix of WSDs. This data matrix is then projected to $\mathbb{R}^{2}$ and $\mathbb{R}^{3}$ using the random projection method described here so that the difference in the structure of these networks can be established using visual inspection of resulting plots.

\subsection{Non-randomness in networks}

Graph non-randomness is a measure proposed to quantify the degree of randomness in a graph [22,23]. Though the notion has been developed initially for social networks, the definitions being formal can be extended to other complex networks. The non-randomness of a graph has been defined for graphs with community structure, where the assumption is that the presence of edges within a community are a result of non-random interactions between components. Though this idea has been used to characterize non-randomness in social networks, we can extend the notion to predator-prey networks as the presence of $3-cycles$ i.e. three given species or group of species from a family sharing a  pairwise mutual carbon-flow relationship through predation among each other may not occur at random. Communities in general showcase a locally cliquish topology.

\subsubsection{Communities in networks}

Communities in a network are set of vertices that have more edges linking the vertices within the set as compared to vertices outside of the set [54]. A partition of the vertex set of a graph is said to form a good community structure if the value of modularity associated with it is close to one. Consider a partition of a network into k communities, represented by a $k\times k$ symmetric matrix $E$ whose element $E_{ij}$ is the fraction of all the edges connecting vertices in community $i$ with vertices in community $j$. Given such a matrix E, modularity is defined as
\begin{equation}
Q = \sum_{i} (E_{ii} - a_{i}^{2}) = Tr(E) - || E^2 ||\;,
\end{equation}
where $a_{i} = \sum_{j} E_{ij}$ represent the fraction of edges with one end in community $i$, $Tr\left(E\right)$ is trace of matrix $E$ and $|| A ||$ is the sum of elements of matrix $A$ [55]. The maximum value possible for Q for any partition is one. Newman further proposed a fast greedy algorithm for community detection which starts with considering each vertex as a community and iteratively merges two communities such that a maximum level of increase in the value of modularity is achieved on merging the two communities [56].

\subsubsection{Graph non-randomness}

For a given graph $G = (V, E)$ with $n$ vertices and $k$ communities, the spectral decomposition of the adjacency matrix $A$ of $G$ is $A = \sum_{i} \lambda_{i} \textbf{x}_{i} \textbf{x}_{i}^{T}$ where $\lambda_{i}$ are eigenvalues of $A$ and $\textbf{x}_{i}$ are the corresponding eigenvectors such that $\lambda_{1} \geq \lambda_{2} \geq \dots \geq \lambda_{n}$. Consider a matrix $\boldsymbol{\alpha}$ in $\mathbb{R}^{n \times k}$ formed by first $k$ eigenvectors ${\textbf{x}}_{i}, 1 \leq i \leq k$ of $A$. Then, the edge non-randomness $R(u,v)$ is defined as 
\begin{equation}
R(u,v) =  \boldsymbol{\alpha_{u}} \boldsymbol{\alpha_{v}^{T}}\;,
\end{equation}
where $\boldsymbol{\alpha_{u}}$ is the row-vector corresponding to row $u$ in $\boldsymbol{\alpha}$. The node non-randomness $R(u)$ of a vertex $u$ in $G$ is defined as 
\begin{equation}
R(u) =  \sum_{v \in N(u)} R(u,v)\;,
\end{equation}
where $N(u)$ is the neighbourhood set of vertex $u$.\\

The non-randomness of a graph denoted $R_{G}$ is defined as 
\begin{equation}
R_{G} =  \sum_{(u,v) \in E} R(u,v)\;.
\end{equation}
The graph non-randomness can be calculated as the sum of first $k$ largest eigenvalues of the adjacency matrix $A$, as given by the following theorem.
\begin{theorem}
The graph non-randomness of a graph $G = (V,E)$ can be calculated as
\begin{equation}
R_{G} = \sum_{(u,v) \in E} R(u,v) = \frac{1}{2} \sum_{u \in V} R(u) = \sum_{i = 1}^{k} \lambda_{i}\;
\end{equation}
\end{theorem}

\subsubsection{Relative non-randomness}

The non-randomness of a graph may quantify how random a graph is but still it may not be possible to compare non-randomness of graphs of different size and order. To overcome this limitation, relative non-randomness is defined in a graph. The relative non-randomness of a graph is obtained by comparing the graph's non-randomness value with the expectation of non-randomness value of all random graphs generated by Erd\H{o}s - R\'enyi random graph model (ER model). Thus relative non-randomness is

\begin{equation}
R^{*}_{G} = \frac{R_{G} - E(R_{ER})}{\sigma(R_{ER})}\;,
\end{equation}

where $E(R_{ER})$ and $\sigma(R_{ER})$ are expectation and standard deviation of graph non-randomness under ER model.

The following theorem helps us estimate the values of expectation and standard deviation of graph non-randomness for ER model.
\begin{theorem}
For a graph $G$ with $n$ vertices and $k$ communities where each community is generated by ER model with parameters  $frac{n}{k}$ and $p$, then graph non-randomness $R_{G}$ has an asymptotically normal distribution with mean $(n - 2k)p + k$ and variance $2kp(1-p)$, where $p = \frac{2km}{n(n-k)}$.
\end{theorem}

The relative non-randomness is thus calculated as

\begin{equation}
R^{*}_{G} = \frac{R_{G} - ((n - 2k)p + k)}{\sqrt{2kp(1-p)}}\;,
\end{equation}

where $p = \frac{2km}{n(n-k)}$. The absolute value of relative non-randomness quantifies whether how different a given graph is from being a random graph.

\section{Results}

A summery of the predator-prey networks originating from aqueous environments studied in this work are presented here in Table 1.

\begin{table}
\caption{Summery of order, size and conectance (edge density) of aqueous predator-prey networks.}
\label{tab:1}       
%
%
\begin{tabular}{p{1cm}p{3cm}p{2.2cm}p{2.2cm}p{2.2cm}p{1.8cm}}
\hline\noalign{\smallskip}
S. No. & Network & number of vertices & number of edges & connectance & reference\\
\noalign{\smallskip}\svhline\noalign{\smallskip}
1 & Estuary 1  & 25 & 44 & 0.1467 & [24]\\
2 & Estuary 2  & 29 & 73 & 0.1798 & [25]\\
3 & Estuary 3  & 27 & 128 & 0.3647 & [26]\\
4 & Estuary 4  & 48 & 221 & 0.1959  & [27]\\
5 & Lake 1  & 20 & 55 & 0.2895 & [28]\\
6 & Lake 2  & 24 & 108 & 0.3913 & [29]\\
7 & Lake 3  & 22 & 77 & 0.3333 & [30]\\
8 & Lake 4  & 50 & 381 & 0.3110 & [31]\\
9 & Marine 1  & 29 & 198 & 0.4877 & [32]\\
10 & Marine 2  & 46 & 131 & 0.1266 & [33]\\
11 & Marine 3  & 80 & 1391 & 0.4402 & [34]\\
12 & Marine 4  & 44 & 400 & 0.4228 & [35]\\
13 & River 1 & 18 & 32 & 0.2092 & [36]\\
14 & River 2  & 29 & 105 & 0.2586 & [37]\\
15 & River 3  & 40 & 180 & 0.2308 & [38]\\
16 & Oxbow lake 1  & 39 & 245 & 0.3306 & [39]\\
17 & Stream 1 & 45 & 193 & 0.1949 & [40]\\
18 & Stream 2  & 24 & 91 & 0.3297 & [41]\\
19 & Stream 3 & 68 & 126 & 0.0553 & [42]\\
20 & Stream 4  & 80 & 155 & 0.0491 & [43]\\
\noalign{\smallskip}\hline\noalign{\smallskip}
\end{tabular}
\vspace*{-12pt}
\end{table}

A similar description of predator-prey networks originating from terrestrial environments is presented in Table 2.

\begin{table}
\caption{Summery of order, size and conectance (edge density) of terrestrial predator-prey networks.}
\label{tab:2}       
%
%
\begin{tabular}{p{1cm}p{3cm}p{2.2cm}p{2.2cm}p{2.2cm}p{1.8cm}}
\hline\noalign{\smallskip}
S. No. & Network & number of vertices & number of edges & connectance & reference\\
\noalign{\smallskip}\svhline\noalign{\smallskip}
1 & Forest 1  & 165 & 114 & 0.0084 & [44]\\
2 & Forest 2  & 35 & 69 & 0.1160 & [45]\\
3 & Forest 3  & 103 & 268 & 0.0510 & [44]\\
4 & Forest 4  & 30 & 66 & 0.1517  & [46]\\
5 & Rocky Shore 1  & 22 & 35 & 0.1515 & [47]\\
6 & Rocky Shore 2  & 27 & 62 & 0.1766 & [48]\\
7 & Rocky Shore 3  & 37 & 79 & 0.1186 & [49]\\
8 & Rocky Shore 4  & 21 & 57 & 0.2714 & [50]\\
9 & Caribbean food-web & 44 & 218 & 0.2304 & [51]\\
10 & Island 1  & 31 & 43 & 0.0925 & [52]\\
11 & Grassland 1  & 133 & 416 & 0.0474 & [53]\\
\noalign{\smallskip}\hline\noalign{\smallskip}
\end{tabular}
\vspace*{-12pt}
\end{table}

Bin selection for WSDs was performed using the method described in section ~\ref{sec:2}. A total of twenty bins were selected for the current study with equal weight of function $w(x) = (1-x)^{4}$ in each bin. Thereafter the weighted spectral distribution for each of the thirty one networks originating from either aqueous environment or terrestrial environment was calculated using the bins. The WSDs were plotted subsequently. The plots of WSDs for aqueous environments and terrestrial environments are given here as Fig. 1 and Fig. 2 respectively.

\begin{figure}[h]
\sidecaption
\includegraphics[scale=.25]{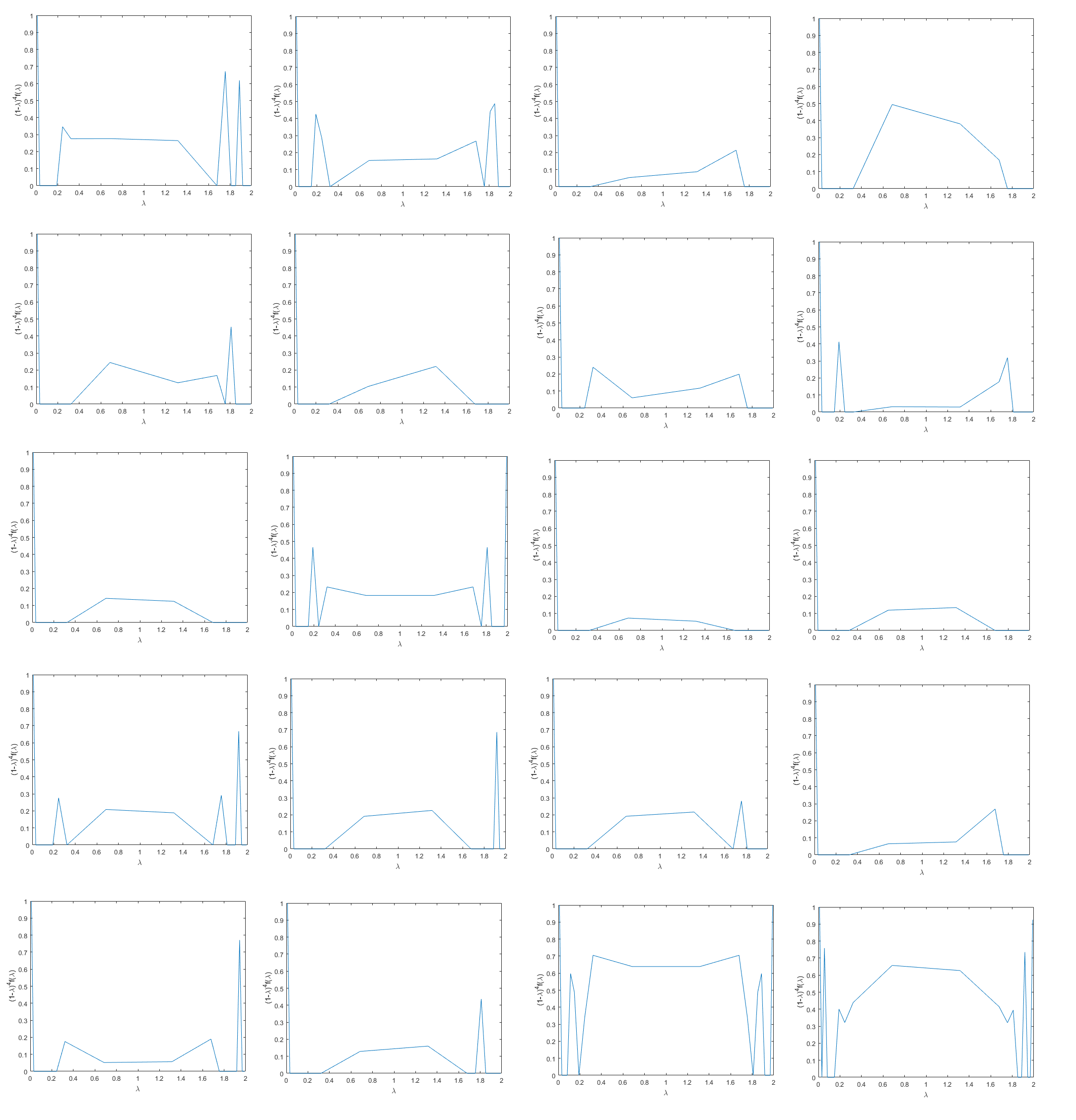}
%
%
\caption{Weighted spectral distributions of aqueous predator-prey networks.}
\label{fig:1}       
\end{figure}

\begin{figure}[h]
\sidecaption
\includegraphics[scale=.25]{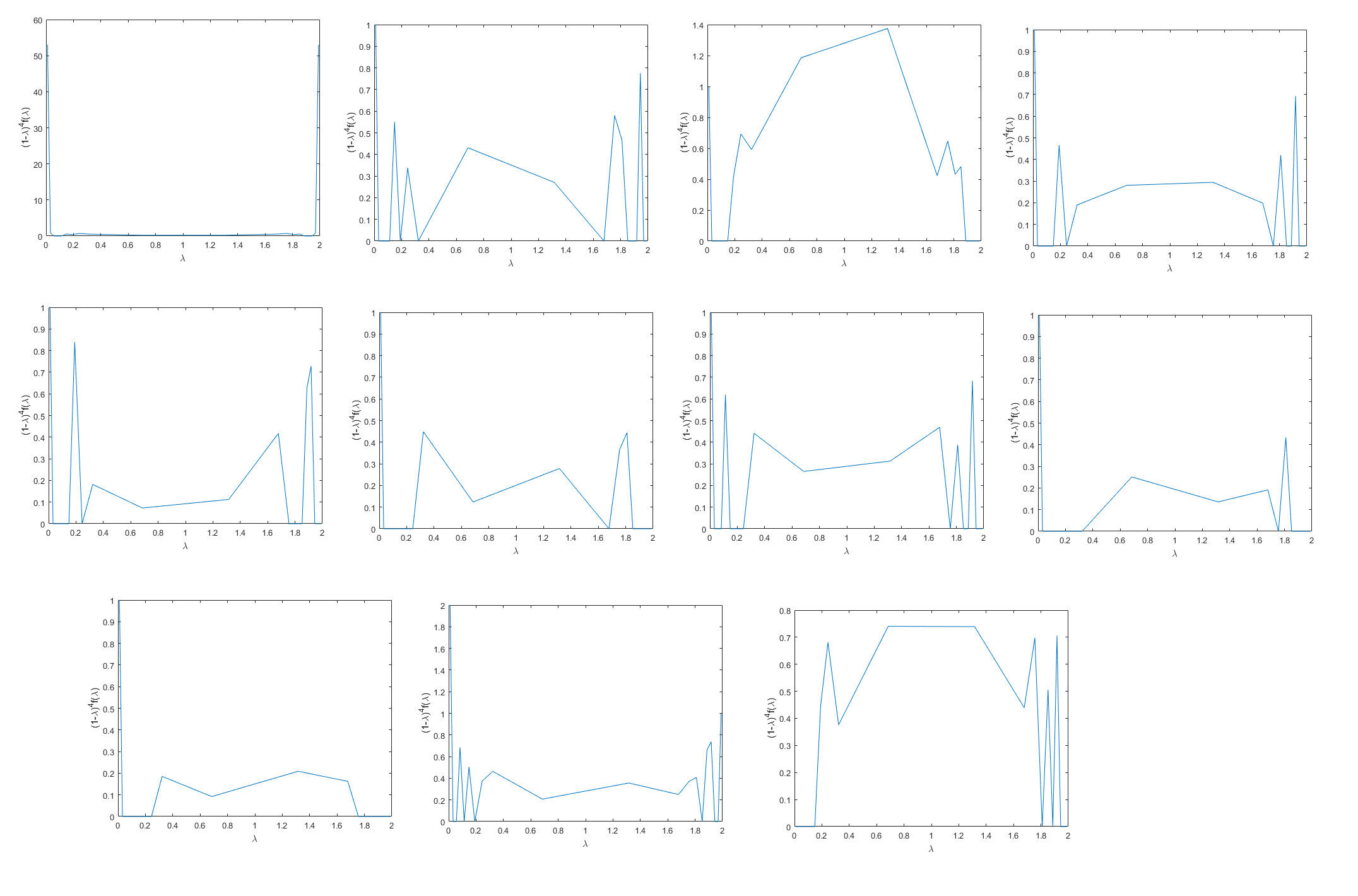}
%
%
\caption{Weighted spectral distributions of terrestrial predator-prey networks.}
\label{fig:2}       
\end{figure}

The WSDs of there networks were projected to $\mathbb{R}^{2}$ and $\mathbb{R}^{3}$ using the random projection method. The plot for random projection of WSD to $\mathbb{R}^{2}$ for all the thirty one networks (including the bipartite networks) is given here as Fig. 3.

\begin{figure}[h]
\sidecaption
\includegraphics[scale=.8]{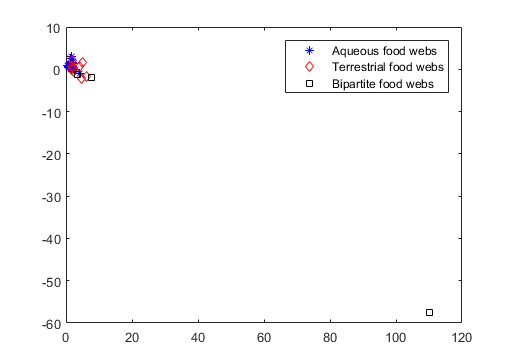}
%
%
\caption{Random projection of weighted spectral distributions of all the networks originating from both the aqueous and terrestrial environments to $\mathbb{R}^{2}$. The networks which are bipartite are marked using black square irrespective of weather they originate from aqueous or terrestrial environment. The axis in this graph is irrelevant, only the separation between the points is of significance.}
\label{fig:3}       
\end{figure}

It is found that one of the bipartite networks studied here (CN) is significantly different from all other networks in terms of the structure as observed by the random projections of WSDs of all the networks to $\mathbb{R}^{2}$. Hence, this network is regarded as an outlier in terms of the structure of networks and thus removed from the study. The remaining thirty networks are projected again to $\mathbb{R}^{2}$ to assess the difference in their structure and the resultant plot is given here as Fig. 4. 

\begin{figure}[h]
\sidecaption
\includegraphics[scale=.8]{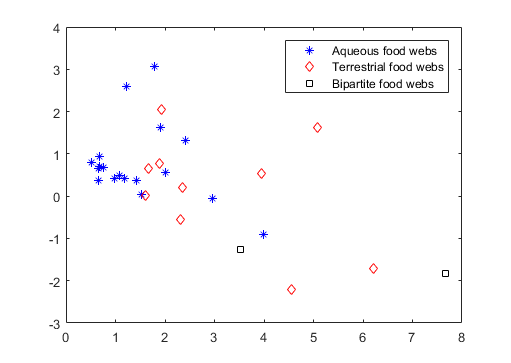}
%
%
\caption{Random projection of weighted spectral distributions networks originating from both the aqueous and terrestrial environments to $\mathbb{R}^{2}$ except the outlier bipartite network seen in Fig. 3. The networks which are bipartite are marked using black square irrespective of weather they originate from aqueous or terrestrial environment. The axis in this graph is irrelevant, only the separation between the points is of significance.}
\label{fig:4}       
\end{figure}

The random projections of WSDs of the thirty remaining networks is plotted to $\mathbb{R}^{3}$ and the resultant plot is given here as Fig. 5.

\begin{figure}[h]
\sidecaption
\includegraphics[scale=.48]{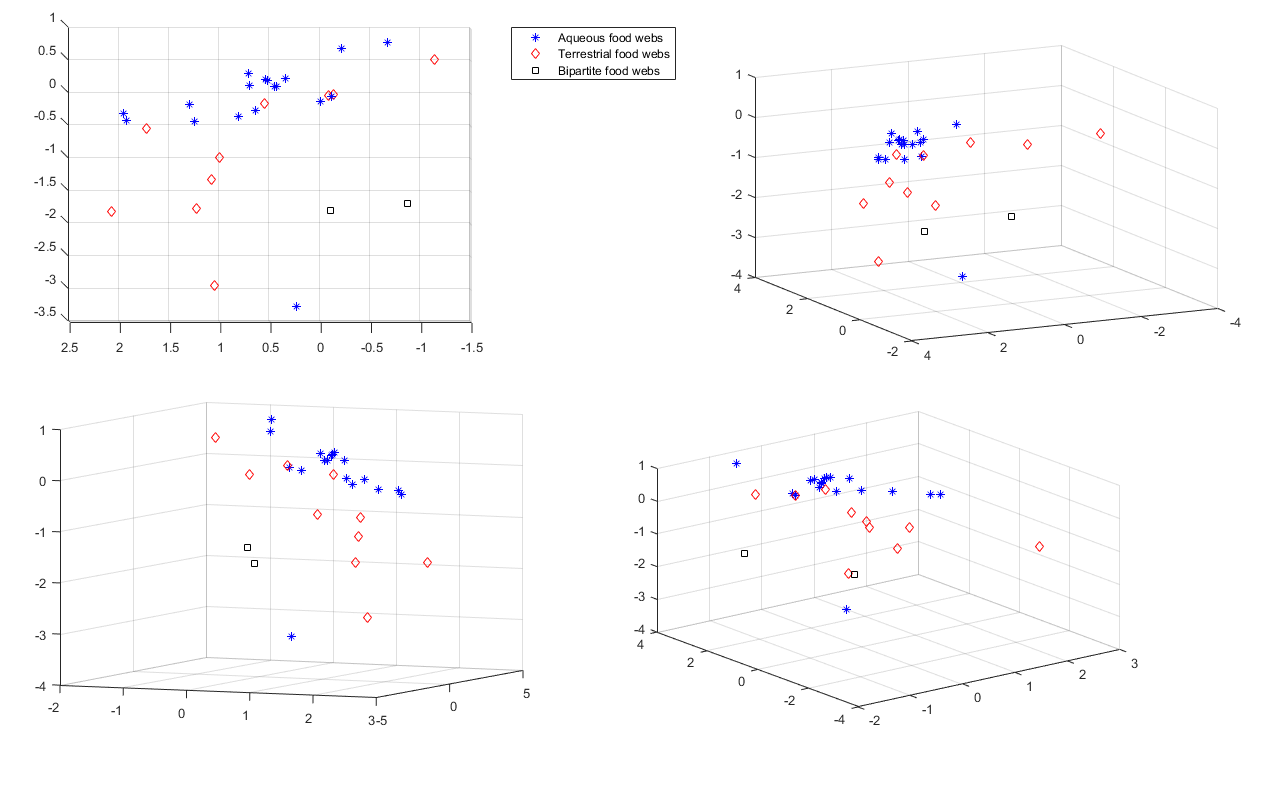}
%
%
\caption{Random projection of weighted spectral distributions networks originating from both the aqueous and terrestrial environments to $\mathbb{R}^{3}$ except the outlier bipartite network seen in Fig. 3. The networks which are bipartite are marked using black square irrespective of weather they originate from aqueous or terrestrial environment. The axis in this graph is irrelevant, only the separation between the points is of significance.}
\label{fig:5}       
\end{figure}

The graph non-randomness and relative non-randomness for each network was calculated subsequently. The results for graph non-randomness and relative non-randomness for networks originating from aqueous environments are presented here as Table. 3. 

\begin{table}
\caption{Summery of graph non-randomness and relative non-randomness of aqueous predator-prey networks.}
\label{tab:3}       
%
%
\begin{tabular}{p{1cm}p{3.2cm}p{3.2cm}p{3.2cm}}
\hline\noalign{\smallskip}
S. No. & Network & graph non-randomness & relative non-randomness\\
\noalign{\smallskip}\svhline\noalign{\smallskip}
1 & Estuary 1  & 11.5955 & 2.8602\\
2 & Estuary 2  & 15.4234 & 4.9062\\
3 & Estuary 3  & 17.8012 & 8.7906\\
4 & Estuary 4  & 29.4796 & 18.6242\\
5 & Lake 1  & 11.9785 & 11.1384\\
6 & Lake 2  & 18.5826 & 4.1888\\
7 & Lake 3  & 13.1922 & 1.5360\\
8 & Lake 4  & 29.3781 & 1.9026\\
9 & Marine 1  & 22.6698 & 7.1029\\
10 & Marine 2  & 18.4632 & 4.3642\\
11 & Marine 3  & 63.8494 & 23.2038\\
12 & Marine 4  & 28.3249 & 12.1853\\
13 & River 1 & 10.3992 & 10.4673\\
14 & River 2  & 11.8438 & 3.5757\\
15 & River 3  & 19.5547 & 7.5896\\
16 & Oxbow lake 1  & 19.9195 & 6.2664\\
17 & Stream 1 & 21.7529 & 12.6408\\
18 & Stream 2  & 12.3089 & 3.7591\\
19 & Stream 3 & 22.0476 & 2.4292\\
20 & Stream 4  & 21.7175 & 2.1053\\
\noalign{\smallskip}\hline\noalign{\smallskip}
\end{tabular}
\vspace*{-12pt}
\end{table}

The graph non-randomness and relative non-randomness for networks originating from terrestrial environments are presented here in Table 4.

\begin{table}
\caption{Summery of graph non-randomness and relative non-randomness of terrestrial predator-prey networks.}
\label{tab:4}       
%
%
\begin{tabular}{p{1cm}p{3.2cm}p{3.2cm}p{3.2cm}}
\hline\noalign{\smallskip}
S. No. & Network & graph non-randomness & relative non-randomness\\
\noalign{\smallskip}\svhline\noalign{\smallskip}
1 & Forest 1  & 72.0897 & 4.3143\\
2 & Forest 2 & 11.5668 & 1.8207\\
3 & Forest 3  & 28.0396 & 1.1832\\
4 & Forest 4  & 13.7766 & 3.8676\\
5 & Rocky Shore 1  & 9.0218 & 1.6464\\
6 & Rocky Shore 2  & 14.0673 & 4.5042\\
7 & Rocky Shore 3  & 13.2562 & 1.1999\\
8 & Rocky Shore 4  & 11.5412 & 6.9958\\
9 & Caribbean food-web  & 22.8054 & 7.0832\\
10 & Island 1  & 13.1662 & 1.9253\\
11 & Grassland 1  & 46.3104 & 4.3532\\
\noalign{\smallskip}\hline\noalign{\smallskip}
\end{tabular}
\vspace*{-12pt}
\end{table}

\section{Discussion and conclusion}

We have examined predator-prey networks originating from different environments and classified the studied networks as either aqueous predator-prey networks or terrestrial predator-prey network depending on whether they originate from aqueous or terrestrial environments respectively. It is often assumed that the structure of food-web networks, in particular predator-prey networks is similar to each other and can be described by some simple algebraic measures such as connectance for instance. 

We calculated the connectance for all the predator-prey networks studied in this work and found that the average value of connectance for the networks originating from aqueous environments is 0.2674 with a standard deviation of 0.1239. The mean and standard deviation of the values of connectance observed for predator-prey networks originating from terrestrial environments is found to be 0.1287 and 0.0792 respectively. Thus on an average the value of connectance observed in predator-prey networks originating from aqueous environments is about double that of the connectance observed for terrestrial predator-prey networks. Thus a difference in the structure of these predator-prey networks is indicated based on the environment from which these networks originate.

The difference observed in the structure of these networks belonging to two different classes (aqueous and terrestrial) is firmly established when we consider observations made by visually inspecting the plots that represent the random projection of WSDs of each network to $\mathbb{R}^{2}$ and $\mathbb{R}^{3}$. It is observed that when the WSDs of all the thirty one networks studied here are randomly projected to $\mathbb{R}^{2}$, one of the network, which is a bipartite graph, has a topology significantly different from the topology of all other networks put together. We mark this network as an outlier in terms of its structure and remove this network from further study. Thereafter, when the WSDs of the remaining thirty networks are again randomly projected to $\mathbb{R}^{2}$, an interesting observation is made.  It is observed that all the aqueous predator-prey networks and terrestrial predator-prey networks are seemingly forming two non-homogeneously overlapping clusters. Since in the plot of random projections of WSDs, as the points closer to each other are similar in terms of their structure, the networks from two classes showing a non-homogeneous overlap clearly indicated the difference in the structure of these networks belonging to the two classes i.e. aqueous and terrestrial. The non-homogeneous overlapping is much significantly observable in random projections of WSDs to $\mathbb{R}^{3}$ and thus our inference, that the structure of networks belonging to the two classes is different, is further supported. 

To further validate that the difference observed in the structure of the predator-prey networks originating from two different environments is due to the precise way in which edges (and thus underlying carbon flow relation between species) are oriented in the networks and is not a result of some property of random graphs, we calculate graph non-randomness and relative non-randomness of each network studied in this work. The average value of relative non-randomness for predator prey-networks originating from aqueous environments and terrestrial environments is found to be equal to 7.4818 and 3.5358 respectively. For a random graph the value of relative non-randomness is close to zero. Thus all the networks studied here show a significant deviation in their structure as compared to the structure of a random network and thus the source of difference observed in the predator-prey networks originating from the two different environments lie in the way edges (and thus consumption relationship) is oriented in these networks.

Thus we conclude that the structure of the predator-prey networks originating from aqueous environments is significantly different from the structure of predator-prey networks originating from terrestrial environments. We further conclude that a universal structural pattern is absent in predator-prey networks originating from different environments. The absence of a universal structural pattern could be a result of difference in the ecological processes in these environments and the observed difference in the structural pattern opens a venue of further research that should be conducted to establish the source of the difference observed in this study. 

We must conclude by asserting that use of properties of graph spectra has been been found to effectively differentiate between the structure of predator-prey networks and thus provide us with an effective tool to study and cross-compare networks originating from different systems.

\end{document}